\documentclass{aa}
\usepackage{graphicx}
\usepackage{natbib}
\usepackage{amsfonts}
\usepackage{amsmath}
\usepackage{amssymb}
\usepackage{amstext}
\begin{document}

   \title{Pseudo Elliptical Lensing Mass Model: 
                application to the NFW mass distribution} 

   \author{Ghislain Golse
          \and
           Jean-Paul Kneib }

   \offprints{Ghislain Golse, \email{ghislain.golse@ast.obs-mip.fr}}

   \institute{Laboratoire d'Astrophysique, Observatoire Midi-Pyr\'en\'ees,
              14 av. E.-Belin, F-31400 Toulouse, France
             }

   \date{Received 18 March 2002 / Accepted 18 April 2002}

   \abstract{We introduce analytical expressions
        for a pseudo fully analytical elliptical projected Navarro, Frenk \& 
        White (NFW) mass profile to be used in lensing equations. 
        We propose a formalism that incorporates
        the ellipticity into the expression for the 
        lens potential, producing a pseudo-elliptical
        mass distribution. This approach can be
        implemented to any circular mass profile for which the projected mass
        profile $\Sigma(r)$ and the deflection angle profile $\alpha(r)$
        both have analytical expressions; however the potential does not
        necessarily need to take an analytical form.
        We apply this new formalism to the NFW mass distribution
        and study how well this pseudo-elliptical NFW model describes
        an elliptical mass distribution. We conclude that
        the pseudo-elliptical NFW model is a good
        description of elliptical mass distributions provided that the
        ellipticity of the projected mass distribution is $\lesssim 0.4$, 
        although with a slightly boxy distribution.

      }

\authorrunning{Golse \& Kneib}
\titlerunning{Pseudo Elliptical NFW Model}

\maketitle

\section{Introduction}

Cosmological N-body simulations of cluster formation \citep{NFW}
indicate the existence of a universal density profile for dark matter
halos, independent of their mass, power spectrum of initial
fluctuations or cosmological parameters. For this so-called NFW
profile, the density increases near the centre with a shallower slope
than an isothermal profile, while it steepens gradually outward and
becomes steeper than isothermal far from the centre. Its analytic
expression is given by
\begin{equation}
\rho(r)=\frac{\rho_{\mathrm c}}{(r/r_s)
(1+r/r_s)^2}
\label{rho_nfw}
\end{equation}
where $\rho_{\mathrm c}$ is a characteristic density and $r_s$ a scale
radius.
Recent higher-resolution simulations \citep[e.g.][]{Moore,Ghigna} advocate
a steeper central cusp of $\rho\propto r^{-1.4}$. Attempts to constrain
the inner slope of the density profile with
high resolution observations of luminosity profiles
\citep{Faber} seems to confirm a central cusp ($\rho\propto r^{-1}$),
rather than a core radius for massive galaxies. On larger scales, 
\citet{Smith} used gravitational lensing to constrain the density profile of
A\,383, a massive galaxy cluster at $z=0.19$, finding a logarithmic slope of
$\sim-1.3$. Robust interpretation of these observational results is complicated
by several factors, including the absence of baryons from high resolution
numerical simulations, systematic uncertainties in the lens models arising from
parametrisation of the mass distribution, and the need to use elliptical mass 
distributions to fit observed multiple image systems.

Gravitational lensing is an ideal tool to constrain the radial structure of 
collapsed halos such as galaxies and clusters of galaxies \citep{Smith}.
However, lensing is only sensitive to the projected mass distribution,
and elliptical mass distributions are needed to match the multiple
images observed in both galaxy and cluster lens systems \citep{Kneib01}.
In response to the debate regarding the inner slope of the density profile,
\citet{Munoz} introduced a general set of ellipsoidal lens
models with $\rho\propto r^{-\gamma}$ as $r \rightarrow 0$ and
$\rho\propto r^{-n}$ at large radius.  
However, as there are  no general analytic expressions for cusped ellipsoidal
mass models, the deflections and magnifications are calculated 
numerically. They applied their model to the gravitational lens 
APM 08279+5255 and found a very
shallow cusp ($\gamma\lesssim0.4$). In contrast, for B~1933+503, they found
that a steep density cusp ($1.6\lesssim\gamma\lesssim2.0$) is favoured.
To avoid expensive numerical integration, \citet{Barkana} suggested an 
alternative solution. For a softened power-law elliptical mass distribution, 
it is possible to approximate the integrand so that the integration can be done
analytically. Therefore, for this flat core model, the deflection can be then 
calculated to high accuracy.

In this paper we propose a new way to introduce ellipticity
in lensing model in a fully analytical way, and we discuss in detail
the recipe and limit of the model for the NFW mass distribution. In 
Sect.~2, we briefly discuss
spherical NFW lens models. Then we present, in Sect.~3, a general 
pseudo-elliptical formalism that incorporates the ellipticity in the 
expression of the lens potential if this is known, or anyway of the
deflection angle. In Sect.~4, we apply this formalism 
to the NFW profile and study the departure of this model from an
elliptical NFW mass model. Finally, in Sect.~5 we discuss prospects 
for the application of this new formalism.

\section{Spherical NFW Lensing Model}

We first recall the expressions for the spherical NFW density profile
\citep[e.g.][]{Bartelmann0,Wright}, this will also allow us to define
all the lensing quantities used hereafter. 

In the thin lens approximation,
we define $z$ as the optical axis and $\Phi(R,z)$ as the three-dimensional
Newtonian gravitational potential -- where $r=\sqrt{R^2+z^2}$. The reduced
two-dimensional lens potential in the plane of the sky is given by 
\citep{Schneider}:
\begin{equation}
\varphi(\vec{\theta})=\frac{2}{c^2}\frac{D_\mathrm{LS}}
{D_\mathrm{OL}D_\mathrm{OS}}\int\limits_{-\infty}^{+\infty} 
\Phi(D_\mathrm{OL}\,\theta,z)\,dz
\label{phi}
\end{equation}

\noindent where $\vec{\theta}=(\theta_1,\theta_2)$ is the angular position
in the image plane.

The deflection angle $\vec{\alpha}$ between the image and 
the source, the convergence $\kappa$ and the shear $\gamma$ are then simply:
\begin{equation}
\left\lbrace
\begin{array}{l}
\vec{\alpha}(\theta)=\vec{\nabla}_{\vec{\theta}}\varphi(\theta)\\
\kappa(\theta)=\displaystyle{\frac{1}{2}\left(\frac{\partial^2\varphi}
{\partial\theta_1^2}
+\frac{\partial^2\varphi}{\partial\theta_2^2}\right)}\\
\gamma^2(\theta)=\|\vec{\gamma}(\theta)\|^2=\displaystyle{
\frac{1}{4}\left(\frac{\partial^2\varphi}{\partial\theta_1^2}
-\frac{\partial^2\varphi}{\partial\theta_2^2}\right)^2+
\left(\frac{\partial^2\varphi}{\partial\theta_1\partial\theta_2}\right)^2}.
\end{array}
\right.
\label{gamma_nfw}
\end{equation}

\noindent For convenience we introduce the
dimensionless radial coordinates $\vec{x}=(x_1,x_2)=\vec{R}/r_s=
\vec{\theta}/\theta_s$ where $\theta_s=r_s/D_\mathrm{OL}$. In the case
of an axially symmetric lens, the relations become simpler, as the
position vector can be replaced by its norm. The surface mass density
then becomes
\begin{equation}
\Sigma(x)=\int\limits_{-\infty}^{+\infty}\rho(r_s\,x,z)dz=2\rho_c r_s F(x)
\label{sigma}
\end{equation}

\noindent with
\begin{equation*}
F(x)=
\begin{cases}
\displaystyle{\frac{1}{x^2-1}\left(1-\frac{1}{\sqrt{1-x^2}}
\mathrm{arcch}
\frac{1}{x}\right)} & (x<1)\\
\displaystyle{\frac{1}{3}} & (x=1)\\
\displaystyle{\frac{1}{x^2-1}\left(1-\frac{1}{\sqrt{x^2-1}}
\arccos{\frac{1}{x}}\right)}
 & (x>1)
\end{cases}
\end{equation*}

\noindent and  the mean surface density inside the dimensionless radius $x$ is 
\begin{equation}
\overline{\Sigma}(x)=\displaystyle{\frac{1}{\pi x^2}\int\limits_{0}^{x}
2\pi x\Sigma(x)dx=4\rho_c r_s \frac{g(x)}{x^2}}
\label{sigma_moy}
\end{equation}

\noindent with
\begin{equation*}
g(x)=
\begin{cases}
\displaystyle{\ln{\frac{x}{2}}+\frac{1}{\sqrt{1-x^2}}\mathrm{arcch}
\frac{1}{x}} & (x<1)\\
\displaystyle{1+\ln{\frac{1}{2}}} & (x=1)\\
\displaystyle{\ln{\frac{x}{2}}+\frac{1}{\sqrt{x^2-1}}\arccos{\frac{1}{x}}}
 & (x>1)
\end{cases}
\end{equation*}

The lensing functions $\vec{\alpha}, \kappa$ and $\gamma$ also have
simple expressions \citep{Miralda0}
\begin{equation}
\left\lbrace
\begin{array}{llcl}
\vec{\alpha}(x)& = & \theta \, 
\displaystyle{\frac{\overline{\Sigma}(x)}{\Sigma_\mathrm{crit}}}
& = 4\kappa_s \, 
\displaystyle{\frac{\theta}{x^2}}g(x)\vec{e}_x \\
\kappa(x) & = & 
\displaystyle{\frac{\Sigma(x)}{\Sigma_\mathrm{crit}}}
& = 2\kappa_s \, F(x)\\
\gamma(x) & = &
\displaystyle{\frac{\overline{\Sigma}(x)-\Sigma(x)}{\Sigma_\mathrm{crit}}}
& = 2\kappa_s \left({\displaystyle{\frac{2g(x)}{x^2}-F(x)}}\right)
\end{array}
\right.
\label{sigma_nfw}
\end{equation}
with $\kappa_s= \rho_c r_s\Sigma_\mathrm{crit}^{-1}$. 
Noting $\vec{\nabla}_{\vec{x}}\,\alpha(x)=
(\partial_{x_1}\alpha,\partial_{x_2}\alpha)$ and $\phi=\arctan(x_2/x_1)$, 
we obtain some useful
relations for the following that hold for any circular mass distribution:
\begin{equation}
\left\lbrace
\begin{array}{lll}
\kappa(x) & = & 
\displaystyle{\frac{1}{2\theta_s}\left(
\frac{\alpha(x)}{x}+\frac{\partial_{x_1}\alpha(\vec{x})}{\cos\phi}\right)}
\\[0.1cm]
\gamma(x) & = &
\displaystyle{\frac{1}{2\theta_s}\left(
\frac{\alpha(x)}{x}-\frac{\partial_{x_1}\alpha(\vec{x})}{\cos\phi}\right)}
\\[0.1cm]
\displaystyle{\frac{\partial_{x_1}\alpha(\vec{x})}{\cos\phi}} & = & 
\displaystyle{\frac{\partial_{x_2}\alpha(\vec{x})}{\sin\phi}}
\\[0.1cm]
\kappa(x)+\gamma(x) & = & \displaystyle{\frac{\alpha(x)}{\theta_s\,x}}
\end{array}
\right.
\label{useful_kg}
\end{equation}

By integrating the
deflection angle, we find the lens potential $\varphi(x)$ \citep{Meneghetti}:
\begin{equation}
\varphi(x)=2\kappa_s\theta_s^2\,h(x)
\label{phi_h}
\end{equation}

\noindent where
\begin{equation}
h(x)=
\begin{cases}
\displaystyle{\ln^2{\frac{x}{2}}-\mathrm{arcch}^2
\frac{1}{x}} & (x<1)\\
\displaystyle{\ln^2{\frac{x}{2}}+\arccos^2{\frac{1}{x}}}
 & (x\ge1)
\end{cases}
\end{equation}

The velocity dispersion $\sigma(r)$ of this potential, computed with the Jeans
equation for an isotropic velocity distribution,
gives an unrealistic central velocity dispersion $\sigma(0)=0$.
In order to compare the pseudo-elliptical NFW potential with other
potentials, we define a scaling parameter $v_c$ (characteristic velocity)
in terms of the parameters of the NFW profile as follows:
\begin{equation}
v_c^2=\frac{8}{3}\mathrm{G}r_s^2\rho_c
\label{sigma_c}
\end{equation}
Using the value of the critical density for closure of the Universe
$\rho_\mathrm{crit}=3H_0^2/8\pi\mathrm{G}$, we find
\begin{eqnarray}
 & & \frac{\rho_c}{\rho_\mathrm{crit}}=\frac{v_c^2}{H_0^2 r_s^2}=
1.8~10^3\,h^{-2}\times
\label{rho_rho_c}\\
 & & \left(\frac{r_s}{150~\mathrm{kpc}}\right)^{-2}
\left(\frac{v_c}{2000~\mathrm{km~s}^{-1}}\right)^2\nonumber
\end{eqnarray}
We showed \citep{Golse1} that a value $v_c=2000$~km~s$^{-1}$ corresponds to
a velocity dispersion $\sigma_0\sim1200$~km~s$^{-1}$ for a \citet{Hjorth} 
model.

\begin{figure*}
\resizebox{\hsize}{!}{
\includegraphics{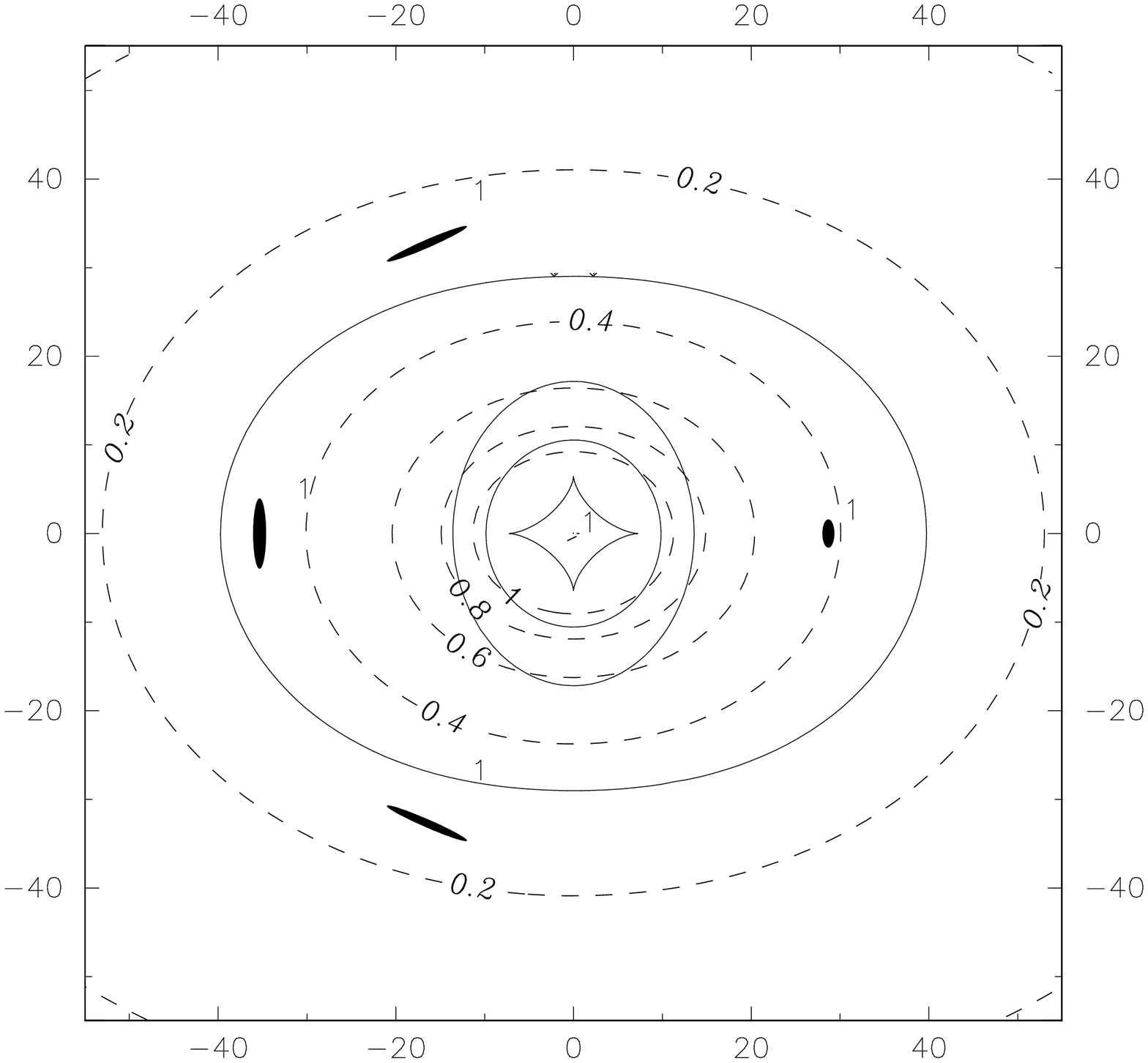}
\hspace*{-5cm}
\includegraphics{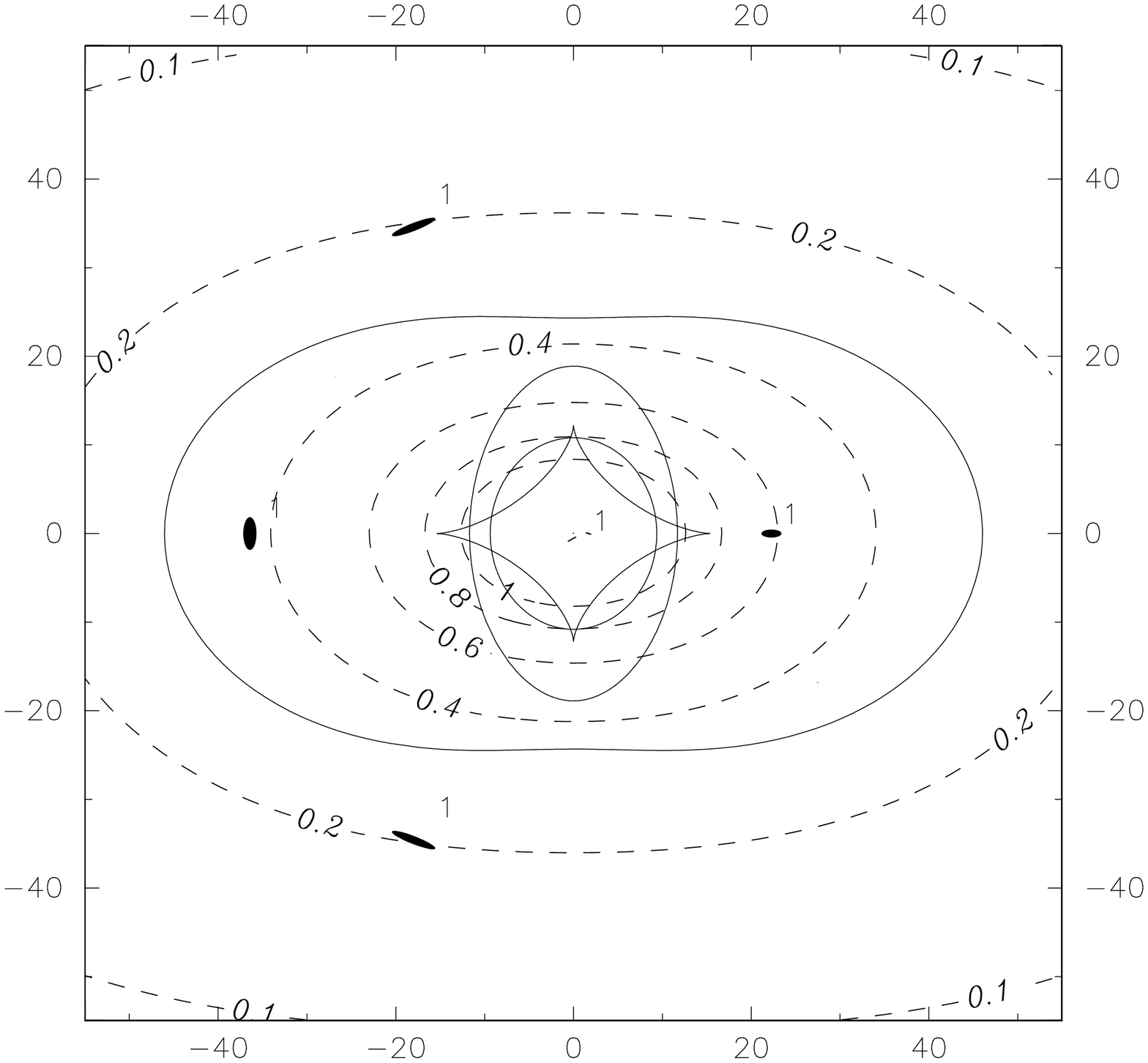}
\hspace*{-5cm}
\includegraphics{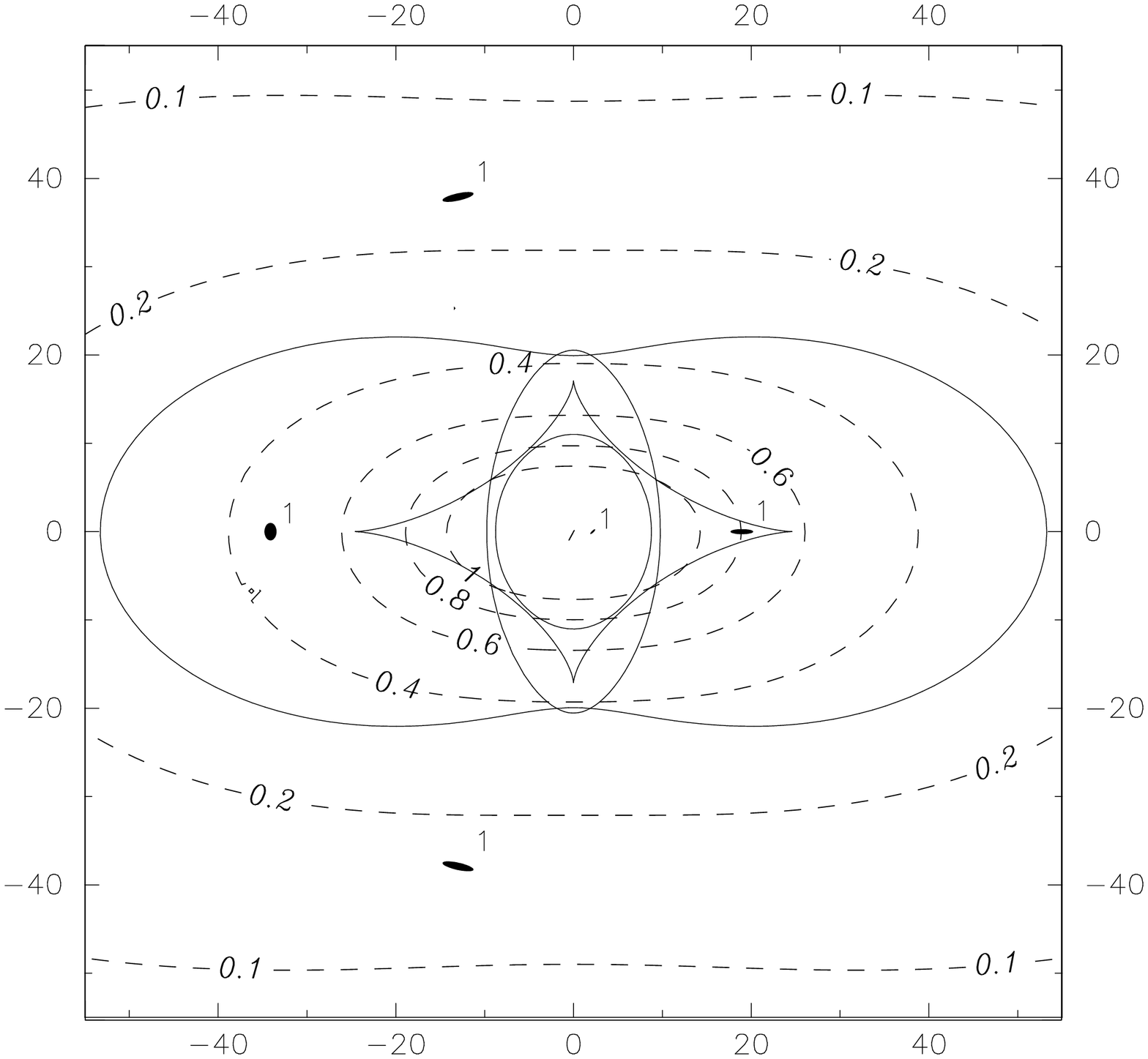}}
\caption{
  System of 5 multiple images generated by a pseudo-elliptical NFW
  cluster at $z_L=0.3$ with the following lens parameters: $v_c=2000$ km/s,
  $\theta_s=31.3$\arcsec ($r_s=150$~kpc) and different values of $\epsilon$. 
  From left to right : $\epsilon=0.1,\ 0.2,\ 0.3$. Solid lines
  are the critical and caustic lines for a source redshift 
  $z_{S1}=1$. Dashed lines are the iso-contours of the 
  dimensionless projected density $\displaystyle{\frac{\Sigma_\epsilon
  (x_1/r_s,x_2/r_s)}{2\rho_c r_s}}$. Units are given in arcseconds.}
\label{clNFW}
\end{figure*}

\section{Elliptical Potential and Deflection-Angle Model}

We will here introduce an ellipticity 
$\epsilon$ in the circular lens potential $\varphi(\theta)$. Moreover,
we assume that the radial profile can be scaled by a scale 
radius $\theta_s$, thus making possible to define $x$ as $x=\theta/\theta_s$
(one can always set $\theta_s=1$ if the radial profile is scale free).
We introduce the ellipticity 
in the expression of the lens potential by substituting $x$ by
$x_\varepsilon$, using the following elliptical coordinate system:
\begin{equation}
\label{defin_ell}
\left\lbrace
\begin{array}{lcl}
x_{1\epsilon} & = & \sqrt{a_{1\epsilon}} \, x_1 \\
x_{2\epsilon} & = & \sqrt{a_{2\epsilon}} \, x_2 \\
x_\epsilon & = & \sqrt{x_{1\epsilon}^2 +
x_{2\epsilon}^2}\ =\  \sqrt{a_{1\epsilon}x_1^2 +a_{2\epsilon}x_2^2}\\
\phi_\epsilon & = & \arctan \left(x_{2\epsilon} / x_{1\epsilon}\right)
\end{array}
\right.
\end{equation}

\noindent where $a_{1\epsilon}$ and $a_{2\epsilon}$ are two parameters defining
the ellipticity. 

Furthermore, from the
elliptical lens potential $\varphi_\epsilon(x)\equiv\varphi(x_\epsilon)$, we 
can then compute the elliptical deviation angle:
\begin{equation}
\vec{\alpha}_\epsilon(\vec{x})=\left(
\begin{array}{l}
\displaystyle{\frac
{\partial\varphi_\epsilon}{\partial x_1}}=
\alpha(x_\epsilon)\sqrt{a_{1\epsilon}}\cos{\phi_\epsilon}\\
\displaystyle{\frac
{\partial\varphi_\epsilon}{\partial x_2}}=
\alpha(x_\epsilon)\sqrt{a_{2\epsilon}}\sin{\phi_\epsilon}\\
\end{array}
\right)
\label{defl_ell}
\end{equation}

We note that these expressions holds for any definition of 
$a_{1\epsilon}$ and $a_{2\epsilon}$.

For instance, \citet{Meneghetti} use for their NFW elliptical
model:
\begin{equation}
\label{a_Meneg}
\begin{array}{ccl}
a_{1\epsilon} & = & 1-\epsilon\\
a_{2\epsilon} & = & 1/(1-\epsilon)
\end{array}
\end{equation}

\noindent for an
ellipticity along the $x_1$ axis. This choice has the advantage to stick to a
'standard' definition: $\epsilon=1-b/a$ --~where $a$ and $b$ are 
respectively the semi-major and semi-minor axis of the elliptical shape. 

However this choice of $a_{1\epsilon}$ and $a_{2\epsilon}$ does
not yield simple expressions for lensing quantities {\it e.g.} $\kappa$ and 
$\gamma$ \citep[see][]{Meneghetti}). 
Nevertheless, we will now show that it is possible to derive simple 
analytic expressions  of  $\kappa$ and
$\gamma$ for a {\it particular} choice of $a_{1\epsilon}$ and $a_{2\epsilon}$. 

At this point, our proposed method can be considered twofold. 
{\it i)} Either the circular lens potential $\varphi$
and the 2D surface mass density $\Sigma$ both have analytic expressions. We can
then introduce the elliptical formalism (\ref{defin_ell}) in the lensing 
potential $\varphi$ and derive the elliptical deflection angle 
$\vec{\alpha}_\epsilon(\vec{x})$ (Eq.~(\ref{defl_ell})). 
{\it ii)} Or, there is no analytic expression for the circular
lens potential (indeed, in many cases the circular lens potential has not 
a simple analytical expression). In this case, we need analytic 
expressions for both the circular deviation angle $\vec{\alpha}$ and the 2D 
surface mass density $\Sigma$. The elliptical formalism (\ref{defin_ell}) 
is then introduced in
the expression of the deflection angle as in Eq.~(\ref{defl_ell}). 
The way the deviation angle is defined ensures that
$\vec{\alpha}(\vec{x})$ derives from a lens potential 
$\varphi_\epsilon(x)\equiv\varphi(x_\epsilon)$, even if there is no analytical
expression for $\varphi(x)$.

Thus, in the following, we will refer to this method as the
elliptical deflection angle model, whether the lens potential is analytically
known or not. To be able to simply derive the
convergence and the shear, we choose the following elliptical parameters:
\begin{equation}
\label{a_GK}
\begin{array}{c}
a_{1\epsilon}=1-\epsilon\\
a_{2\epsilon}=1+\epsilon
\end{array}
\end{equation}

\noindent For small $\epsilon$, it gives the same ellipticity along the $x_1$
axis as the one given by the parameters defined in Eqs~(\ref{a_Meneg}). More
generally, if we denote by $\epsilon_\varphi$ the ellipticity of the lens
potential contours --~taken as $1-b/a$~--, we have:
\begin{equation}
\epsilon_\varphi=1-\displaystyle{\sqrt{\frac{1-\epsilon}{1+\epsilon}}}
\label{eps_phi}
\end{equation}

\noindent independently of the profile. This means there is a direct
relation between the 'standard' ellipticity and the deflection 
angle ellipticity.

However, for this particular choice of  $\epsilon$ we can derive 
easily --~using Eqs~(\ref{useful_kg})~-- the corresponding convergence 
$\kappa_\epsilon(\vec{x})$
induced by Eq.~(\ref{defl_ell}):
\begin{eqnarray}
\kappa_\epsilon(\vec{x}) & = & \displaystyle{\frac{1}{2\theta_s^2}\left(
\frac{\partial^2\varphi_\epsilon}{\partial x_1^2}
+\frac{\partial^2\varphi_\epsilon}{\partial x_2^2}\right)} \nonumber \\
 & = & 
\displaystyle{\kappa(\vec{x}_\epsilon)+\frac{\epsilon}{2\theta_s^2}\left(
\frac{\partial^2\varphi(x_\epsilon)}{\partial x_{2\epsilon}^2}
-\frac{\partial^2\varphi(x_\epsilon)}{\partial
  x_{1\epsilon}^2}\right)}
\nonumber \\
 & = & \displaystyle{\kappa(\vec{x}_\epsilon)+
\epsilon\cos{2\phi_\epsilon}\,\gamma(\vec{x}_\epsilon)}.
\label{kappa_ell}
\end{eqnarray}

Similarly, the shear $\gamma_\epsilon(\vec{x})$ can be written as:
\begin{equation}
\begin{array}{rl}
\gamma_\epsilon^2(\vec{x})
 & = \displaystyle{\frac{1}{4\theta_s^4}\left\lbrace\left(
\frac{\partial^2\varphi_\epsilon}{\partial x_1^2}
-\frac{\partial^2\varphi_\epsilon}{\partial x_2^2}\right)^2+
\left(2\frac{\partial^2\varphi_\epsilon}{\partial x_1
\partial x_2}\right)^2\right\rbrace}\\
& = \gamma^2(\vec{x}_\epsilon) \\
+ \displaystyle{\frac{\epsilon}{2\theta_s^4}}
& \displaystyle{\left(
\frac{\partial^2\varphi(x_\epsilon)}{\partial x_{2\epsilon}^2}
-\frac{\partial^2\varphi(x_\epsilon)}{\partial x_{1\epsilon}^2}\right)
\left(\frac{\partial^2\varphi(x_\epsilon)}{\partial x_{1\epsilon}^2}
+\frac{\partial^2\varphi(x_\epsilon)}{\partial x_{2\epsilon}^2}\right)}\\
+ \displaystyle{\frac{\epsilon^2}{4\theta_s^4}}
& \displaystyle{\left\lbrace\left(
\frac{\partial^2\varphi(x_\epsilon)}{\partial x_{1\epsilon}^2}
+\frac{\partial^2\varphi(x_\epsilon)}{\partial x_{2\epsilon}^2}\right)^2
-4\left(\frac{\partial^2\varphi(x_\epsilon)}
{\partial x_{1\epsilon}\partial x_{2\epsilon}}\right)^2\right\rbrace}
\end{array}
\end{equation}
which, using Eqs~(\ref{useful_kg}), can be simplified as:

\begin{eqnarray}
\gamma_\epsilon^2(\vec{x})=\gamma^2(\vec{x}_\epsilon) & + & 2\epsilon
\cos{2\phi_\epsilon}\gamma(\vec{x}_\epsilon)\kappa(\vec{x}_\epsilon) 
\nonumber\\
 & + & \epsilon^2(\kappa^2(\vec{x}_\epsilon)-\cos^2{2\phi_\epsilon}
\gamma^2(\vec{x}_\epsilon)).
\end{eqnarray}

Finally, the projected mass density $\Sigma_\epsilon(\vec{x})$ is simply
determined from equations (\ref{kappa_ell}) and (\ref{sigma_nfw}):
\begin{equation}
\Sigma_\epsilon(\vec{x})=\Sigma(\vec{x}_\epsilon)+\epsilon\cos{2\phi_\epsilon}
(\overline{\Sigma}(\vec{x}_\epsilon)-\Sigma(\vec{x}_\epsilon)).
\label{sigma_ell}
\end{equation}

\label{model}

\section{Application of the Model to NFW Halos}

Now, we apply the elliptical deflection angle model developed in
Sect.~\ref{model} to the NFW profile (\ref{rho_nfw}). In that case, 
the lens potential (\ref{phi}) and the 2D projected mass profile (\ref{sigma})
are known analytically.

An illustration of some lensed images using our new formalism
applied to the NFW profile is shown in Fig.~\ref{clNFW}. The
caustic associated with the tangential critical line has the 
usual diamond shape and is not reduced to a central point as in the
spherical NFW case. This of course makes the
formation of 5-image configurations with tangential images possible. 

\begin{figure}
  \resizebox{\hsize}{!}{\includegraphics{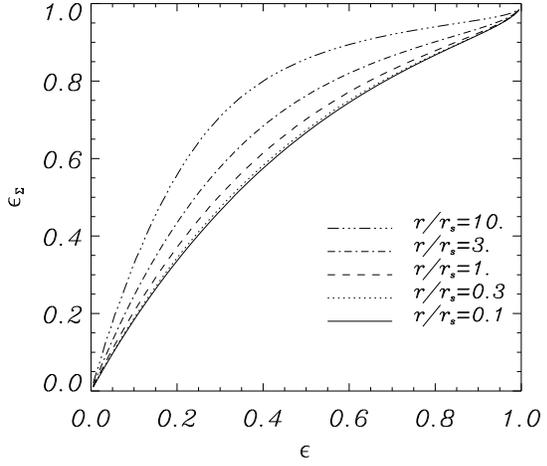}}
\caption{
Ellipticity $\epsilon_\Sigma$ of the projected density $\Sigma_\epsilon$ as a
function of the ellipticity $\epsilon$ defined in 
Eq.(\ref{defin_ell}) with the choice (\ref{a_GK}) for the NFW profile. 
We show curves for different values 
of $r/r_s$ ($r$: ellipse semi diagonal, $r_s$: NFW scale radius).}
\label{epsS}
\end{figure}

\begin{figure}
  \resizebox{\hsize}{!}{\includegraphics{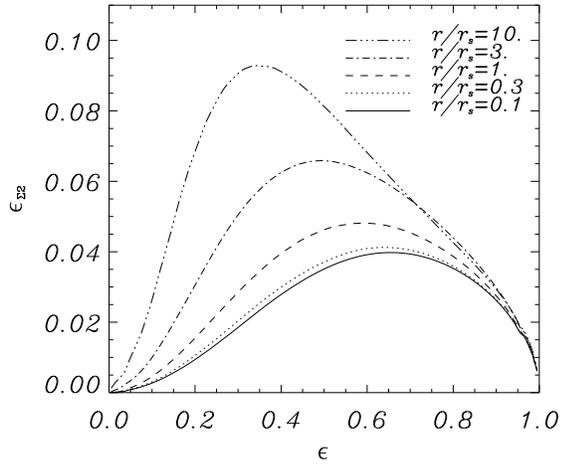}}
\caption{
Parameter $\epsilon_{\Sigma2}$ from Eq.~(\ref{r_theta2}) used to fit the
projected density $\Sigma_\epsilon$; it characterises the deviation from a
purely elliptical curve. $\epsilon_{\Sigma2}$ is a
function of the ellipticity $\epsilon$ defined in 
Eq.(\ref{defin_ell}) with the choice (\ref{a_GK}) for the NFW profile. 
Curves are shown for different values 
of $r/r_s$ ($r$: ellipse semi diagonal, $r_s$: NFW scale radius).}
\label{epsS2}
\end{figure}

\subsection{Expression of the 3D pseudo-elliptical mass distribution}

This particular mass distribution has the advantage that the 3D
pseudo-elliptical NFW mass profile $\rho_\epsilon(\vec{x},x_3)$ can also be
derived. Indeed using the scaled variables $x_3=z/r_s$ and 
$u=r/r_s=\sqrt{x^2+x_3^2}$, we can compute from equations (\ref{sigma}), 
(\ref{sigma_moy}) and (\ref{sigma_ell}):
\begin{eqnarray}
\rho_\epsilon(\vec{x},x_3) & = & \rho(\vec{x}_\epsilon,x_3) \nonumber \\
 & + &\epsilon
\cos{2\phi_\epsilon}\left(\frac{2}{x_\epsilon^2}\int_0^{x_\epsilon}
x\rho(\vec{x},x_3)dx - \rho(\vec{x}_\epsilon,x_3)\right) \nonumber
\\
& = & \rho(\vec{x}_\epsilon,x_3) \nonumber \\
 & + &\epsilon
\cos{2\phi_\epsilon}\left(\overline{\rho}(\vec{x}_\epsilon,x_3) 
- \rho(\vec{x}_\epsilon,x_3)\right)
\label{rho_ell}
\end{eqnarray}

with
\begin{equation}
\overline{\rho}\left(\sqrt{u^2-x_3^2},x_3\right) =
\frac{2\rho_c}{(1+u)(1+x_3)(u+x_3)}
\end{equation}

\subsection{Physical limits of the NFW pseudo-elliptical mass model}

We now investigate the range of
$\epsilon$ for which this NFW mass model is an adequate description of an 
elliptical underlying mass distribution. We will use two methods to quantify
the deviation of our model from a purely elliptical distribution.

Fig.~\ref{clNFW} shows the contours (dashed lines)
of the projected mass density $\Sigma_\epsilon$ (Eq.~(\ref{sigma_ell})) 
for $\epsilon=0.1, 0.2, 0.3$. In the more elliptical models, the contours 
become increasingly boxy/peanut shaped at larger ``radius''. 
In order to investigate this boxy behaviour, we must first quantify the
ellipticity $\epsilon_\Sigma$ of the projected mass distribution 
$\Sigma_\epsilon$, and then relate this to the ellipticity $\epsilon$ of the 
lens model. Purely elliptical projected mass density contours would have a
polar equation of the type
\begin{equation}
r\propto\displaystyle{\frac{1}{\sqrt{(1-\epsilon_\Sigma)\cos^2\phi
+\displaystyle{\frac{\sin^2\phi}{1-\epsilon_\Sigma}}}}}.
\end{equation}

We propose a fit of elliptical-like functions which is a deviation from an
elliptical model. It is slightly different from the function presented by
\cite{Jedrzejewski} and already used by \cite{Shaw} or \citet{Quillen}.
We write the polar equation 
\begin{equation}
r\propto\displaystyle{\frac{1}{\sqrt{(1-\epsilon_\Sigma)\cos^2\phi
+\displaystyle{\frac{\sin^2\phi}{1-\epsilon_\Sigma}}
+\epsilon_{\Sigma2}\cos(4\phi_{\epsilon_\Sigma})}}}
\label{r_theta2}
\end{equation}

\noindent with
\begin{equation}
\phi_{\epsilon_\Sigma}=\arctan\left(
\displaystyle{\frac{\tan\phi}{1-\epsilon_\Sigma}}\right).
\label{theta_eps}
\end{equation}

\noindent If we assume that the fitted contour is roughly an ellipse
with an ellipticity $\epsilon_\Sigma$,
the angular direction $\phi_{ab}$ of its diagonal is such that
$\tan\phi_{ab}=b/a\simeq 1-\epsilon_\Sigma$. In this expression, $a$ and $b$
 are defined by $\Sigma_\epsilon (a,0)=\Sigma_\epsilon(0,b)$, i.e. the 
``pseudo'' semi major and minor axis. Taking a radial coordinate of the form
(\ref{r_theta2}) permits to quantify the degree of boxiness for a 
non elliptical model. Indeed, compared to an ellipse, $r$ is then 
smaller along the axis and larger along the diagonals for 
$\epsilon_{\Sigma2}>0$, i.e. the distribution is boxier. This kind of fit can 
be generally applied
to check quantitatively the deviation of a boxy/peanut function from an
ellipse via the parameter $\epsilon_{\Sigma2}$.

\begin{figure*}
\resizebox{\hsize}{!}{
\includegraphics{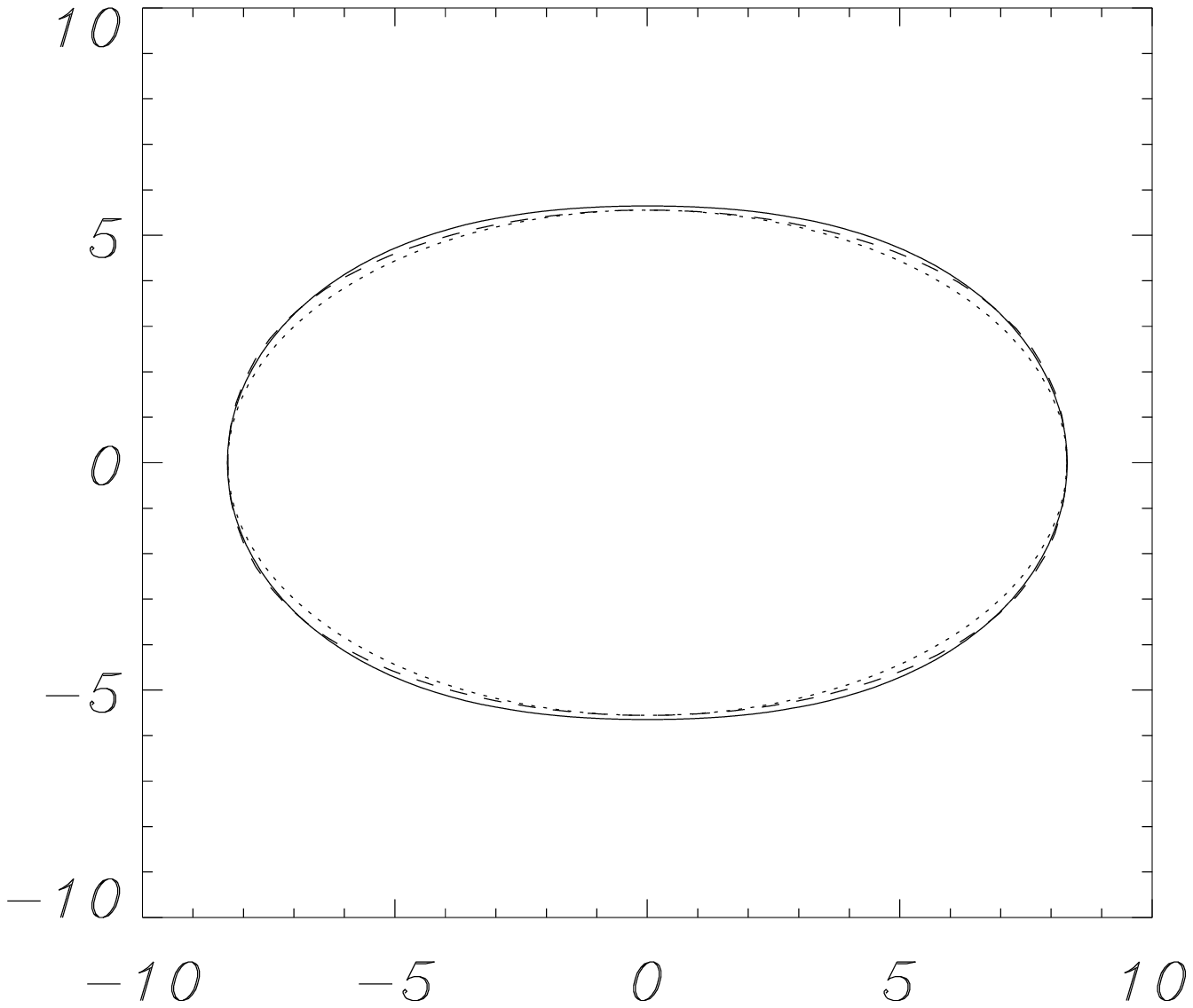}
\hspace*{-4cm}
\includegraphics{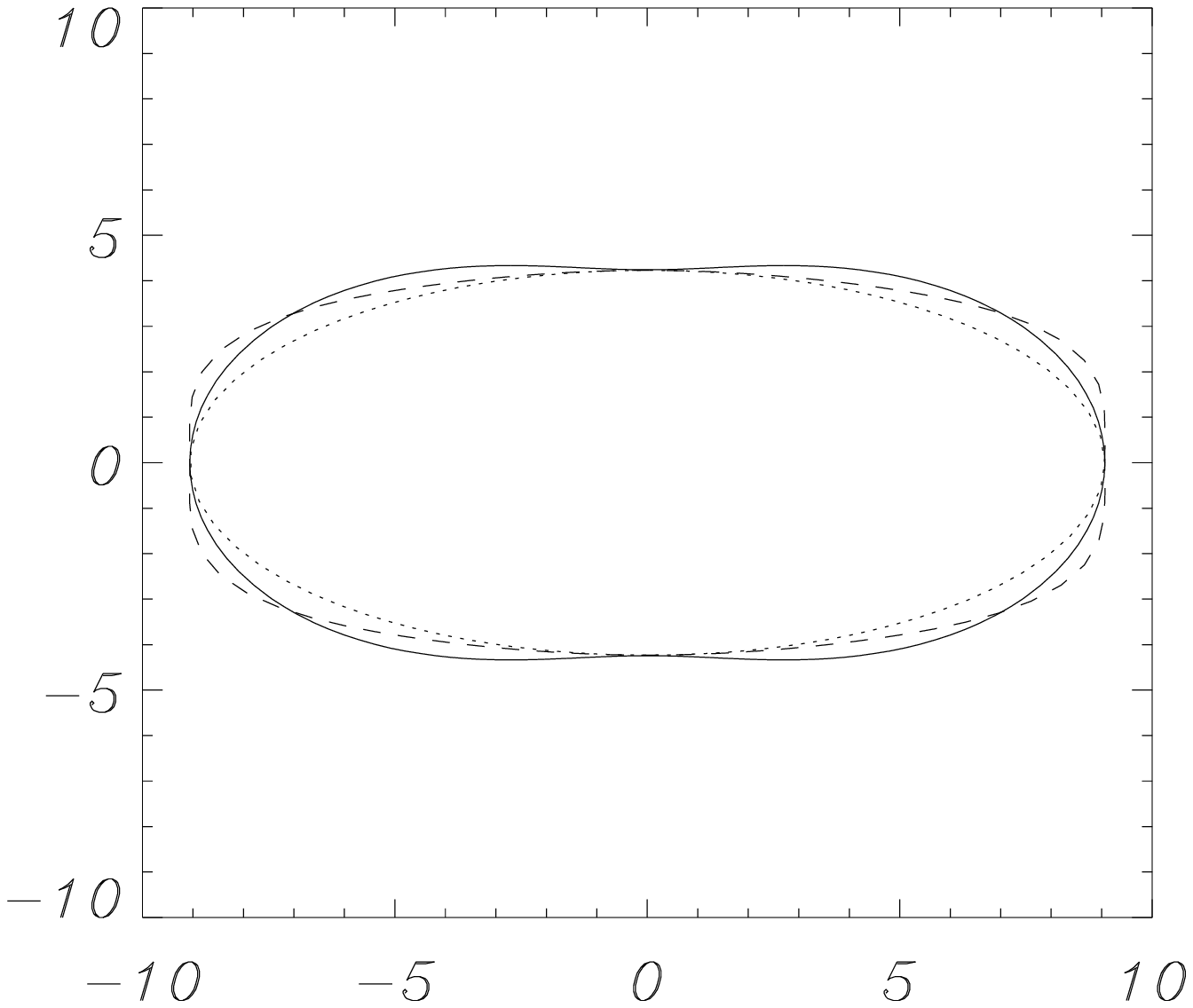}
\hspace*{-4cm}
\includegraphics{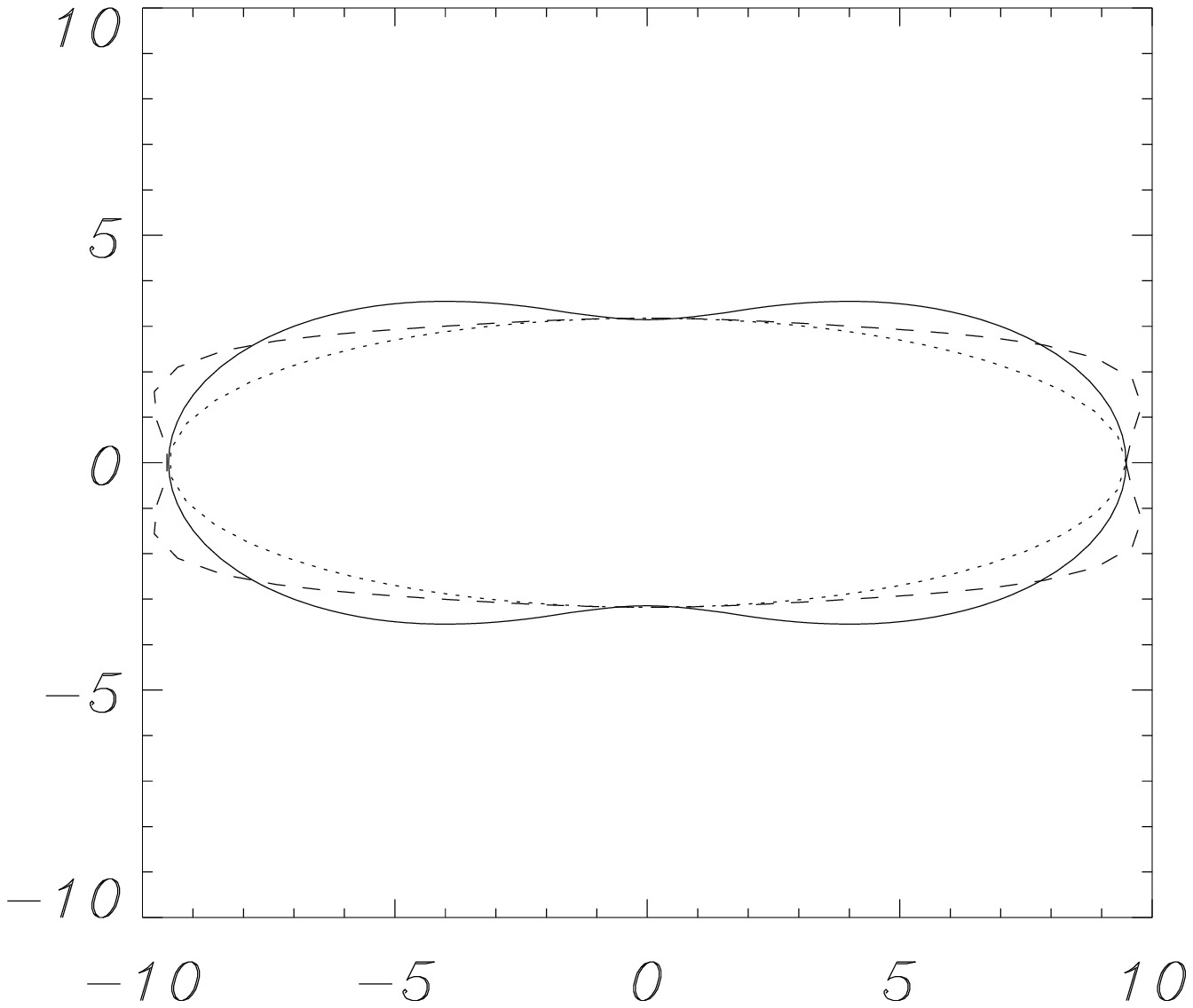}}
\caption{
Solid lines: projected density $\Sigma_\epsilon(x_1/r_s,x_2/r_s)$ contour for
a pseudo-elliptical NFW profile at $r/r_s=10$ ($r$: ellipse semi diagonal,
$r_s$: NFW scale radius). Dotted lines: best fit ellipse. Dashed lines: fitting
function (\ref{r_theta2}) with computed parameters $\epsilon_\Sigma$ and 
$\epsilon_{\Sigma2}$. From left to right: $\epsilon=0.1,0.2,0.3$.}
\label{fit}
\end{figure*}

\begin{figure}
  \resizebox{\hsize}{!}{\includegraphics{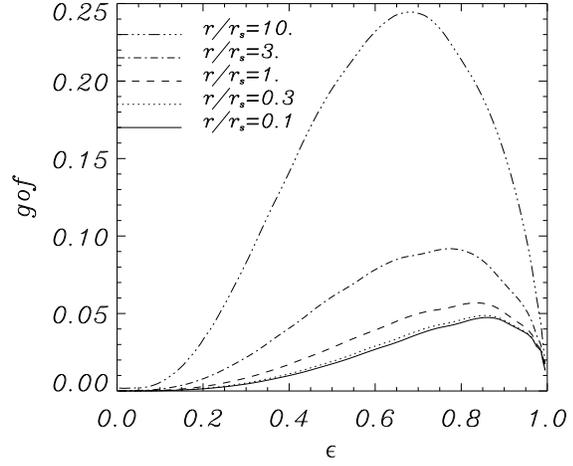}}
\caption{
Goodness of fit of the projected density $\Sigma_\epsilon$ contours by a
function of the type (\ref{r_theta2}). The ellipticity $\epsilon$ is 
defined for 
the NFW profile in Eq.(\ref{defin_ell}) with the choice (\ref{a_GK}). 
Curves are shown for 
different values of $r/r_s$ ($r$: ellipse semi diagonal, $r_s$: NFW scale 
radius).The goodness of fit is computed as in Eq.~(\ref{gof}) with $N=20$.}
\label{gofig}
\end{figure}

For a given ellipticity $\epsilon$, introduced in the deflection angle, and a 
given radius $r=\sqrt{a^2+b^2}$, we will fit the parameters 
$\epsilon_\Sigma$ and $\epsilon_{\Sigma2}$ in the corresponding surface
density profile. 
A goodness of fit indicator will allow us to check how effective 
the representation is. The ratio $b/a$ gives
a first relation. The other one is given by $c/a$ where $c$ is such that
$\Sigma_\epsilon (c,c\tan\phi_{ab})=\Sigma_\epsilon (a,0)=
\Sigma_\epsilon(0,b)$. This means that we adjust the coefficients 
of the fitting function along the first order ellipse diagonal. 
Eq.~(\ref{r_theta2}) is indeed a deviation from an ellipse in this direction.
Actually, getting two relations does not lead analytically to $\epsilon_\Sigma$
and $\epsilon_{\Sigma2}$, mainly because of the angle 
$\phi_{\epsilon_\Sigma}$ which depends on $\epsilon_\Sigma$. So we assume
in practice that $1-\epsilon_\Sigma\simeq b/a$. This approximation is correct
since
\begin{equation}
\frac{b}{a}=\displaystyle{\frac{\sqrt{1-\epsilon_\Sigma+\epsilon_{\Sigma2}}}
{\sqrt{\displaystyle{\frac{1}{1-\epsilon_\Sigma}}+\epsilon_{\Sigma2}}}}
\simeq(1-\epsilon_\Sigma)\left(1+\frac{\epsilon_\Sigma\epsilon_{\Sigma2}}{2}
\left(\frac{2-\epsilon_\Sigma}{1-\epsilon_\Sigma}\right)\right)
\end{equation}

\noindent for $\epsilon_{\Sigma2}\ll 1$ and $1-\epsilon_\Sigma=O(1)$. It is 
then possible to express $\epsilon_\Sigma$ and 
$\epsilon_{\Sigma2}$ analytically for given $\epsilon$ and $r$, see
Figs~(\ref{epsS}) and (\ref{epsS2}).

We note that a given value of $\epsilon$ corresponds to
a higher value of $\epsilon_\Sigma$ (Fig.~(\ref{epsS})). $\epsilon$ can be
considered as the
ellipticity of the potential $\epsilon_\varphi$ for a large range of values
(see Eq.~(\ref{eps_phi}): there is less than 10\% error for $\epsilon\le0.25$).
It is also known that the ellipticity of the 
projected mass density is proportional to and larger than the ellipticity of 
the potential in the linear approximation and then flattens \citep{Kneib}. For
instance, a singular isothermal ellipse satisfies
$\epsilon_\Sigma=3\,\epsilon_\varphi$ for $\epsilon_\varphi\ll 1$. 

To derive numerically such a relation for the NFW profile, we need to know the
range of acceptable and physical values for $\epsilon$. For all ellipticities 
and up to $r/r_s=10$ (i.e. $r\sim 1.5$~Mpc for a galaxy cluster), 
$\epsilon_{\Sigma2}<0.1$ (see Fig.~(\ref{epsS2})). So the deviation parameter 
is not too large and the elliptical approximation could be considered as 
acceptable if the goodness of fit for the function (\ref{r_theta2}) is small.
To check the relevance of this fit, we plot the $\Sigma_\epsilon$ contour, the
first order ellipse and the fitting function found for $r/r_s=10$ and 
$\epsilon=0.1,0.2,0.3$ (Fig.~(\ref{fit})). The fit is correct for small
ellipticities but is not suited for $\epsilon=0.3$. In particular it fails to 
reproduce the shape along the $x_2$ axis.

Quantitatively, we define a goodness of fit in the following way:
\begin{equation}
gof=\frac{1}{N}\sum_{i=1}^{N}\frac{|r_\Sigma(\phi_i)-r(\phi_i)|}
{r(\phi_i)}.
\label{gof}
\end{equation}

\noindent for $\phi_i=\displaystyle{\frac{\pi}{2}\frac{i}{N}}$
(for symmetry reasons). For given
$a$ and $\epsilon$, $r_\Sigma(\phi_i)$ and $r(\phi_i)$ are respectively 
the distances of the projected 
density contour and of the corresponding fit function (\ref{r_theta2}) from 
the centre in the direction $\phi_i$. Fig.~(\ref{gofig}) confirms that the 
goodness of fit becomes quite large from $\epsilon\sim0.25$ (the deviation 
from the proposed function then reaches 10\%).

We think that function (\ref{r_theta2}) can be useful to test deviation from
ellipticity of a given function in various sets of problems. In our case the
deviation parameter $\epsilon_{\Sigma2}$ is rather small but 
the goodness of fit is only acceptable for
ellipticities (introduced in the deviation angle) of $\epsilon<0.25$.

\begin{figure}
  \resizebox{\hsize}{!}{\includegraphics{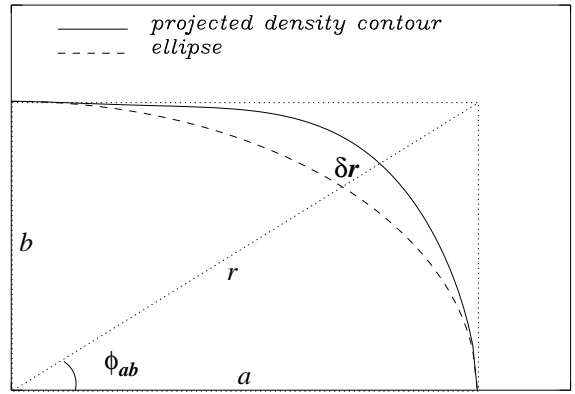}}
\caption{
Method used to compare a projected density contour and a real ellipse with
semi axes $a$ and $b$. $\delta r/r$ characterises this deviation.
}
\label{ellipse}
\end{figure}

\begin{figure}
  \resizebox{\hsize}{!}{\includegraphics{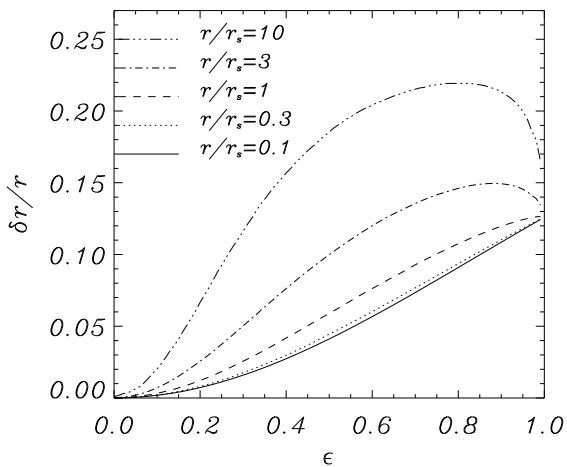}}
\caption{
$\delta r/r$ (as defined in Fig.~\ref{ellipse}) as a function of $\epsilon$.
It characterises the deviation of the projected density from an ellipsoidal
model for various $r/r_s$ ratios ($r_s$: NFW scale radius).
}
\label{dr_ell}
\end{figure}

Alternatively, to simply quantify the degree of boxiness 
for this pseudo-elliptical NFW model, we defined the characteristic deviation 
from ellipticity in the following way.
On Fig.~\ref{ellipse}, $\delta r$ is the distance between a real ellipse and
a $\Sigma_\epsilon$ contour along the ellipse diagonal. We plot $\delta r/r$
versus $\epsilon$ for different $r/r_s$ ratios in Fig.~\ref{dr_ell}.
At all radii, and for all $\epsilon$, the model has a positive $\delta r$,
i.e. the model mass distribution is more boxy than an elliptical distribution.
Assuming that the underlying mass distribution is elliptical, and aiming
to incur an error in $r$ which is $\lesssim10$\%, we find that on scales of 
1.5 Mpc (i.e. corresponding to $r/r_s\sim10$ for a galaxy cluster), the 
pseudo-elliptical model provides an adequate description of the underlying mass
distribution for $\epsilon\lesssim0.25$, which translates to a limit of
$\epsilon_\Sigma\lesssim0.4$ on the projected density at $r/r_s=1$ 
(see Fig.~\ref{epsS}).

\begin{figure}
  \resizebox{\hsize}{!}{\includegraphics{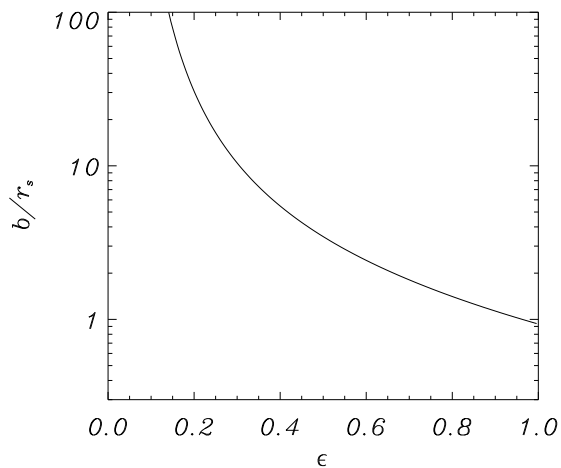}}
\caption{
$b/r_s$ ($b$: distance from the centre along the $x_2$ axis at which 
$\Sigma_\epsilon$ becomes negative, $r_s$: NFW scale radius) as a function of 
$\epsilon$.}
\label{b_ell}
\end{figure}

For models in which the potential -- rather than the 
deflection angle -- is chosen to have elliptical contours, the corresponding
density contours acquire the artificial feature of a dumbbell shape, and the
density can also become negative \citep{Kassiola}. Similarly here, for large 
ellipticities or at large radii, we see from Eq.(\ref{sigma_ell})
that the projected density $\Sigma_\epsilon$ can also become negative. 
This occurs
closer to the centre along the $x_2$ axis where $\cos{2\phi_\epsilon}=-1$. For
each value of the ellipticity $\epsilon$, we plot in Fig.~\ref{b_ell} the 
scaled distance $b/r_s$ at which $\Sigma_\epsilon(0,b)$ becomes negative. 
If we decide to have physical (i.e. positive) mass density up a scale 
of 1.5 Mpc (typically $b/r_s=10$ for a cluster), we have to restrict ourselves
to ellipticities smaller than $\epsilon\sim0.3$ (i.e. $\epsilon_\Sigma\sim0.6$
at $r/r_s=1$ from Fig.~\ref{epsS}). So a relatively broad 
range of systems can be modelled in a physically consistent way.

We want to obtain an explicit, if approximate, relationship 
between the ellipticity $\epsilon$ introduced in the deviation angle and the
ellipticity of the projected mass density $\epsilon_\Sigma$ it induces.
In the acceptable and physical range $[0,0.25]$ for $\epsilon$, we fit
a polynomial of the form:
\begin{equation}
\epsilon_\Sigma=a_1\,\epsilon+a_2\,\epsilon^2.
\end{equation}
A fit for $r=r_s$ leads to $\epsilon_\Sigma=2.27\,\epsilon
-2.03\,\epsilon^2$ with a $\chi^2=3.9\,10^{-7}$.
More generally,  the coefficients $a_i$ depend on $x=r/r_s$. A fit 
between $x=0$ and $x=10$ gives

\begin{equation}
\left\lbrace
\begin{array}{ccrcc}
a_1 & = & 2.12 & + & 0.179\,x\\
a_2 & = & -1.70 & - & 0.328\,x
\end{array}
\right.
\label{a_i}
\end{equation}

\noindent with a $\chi^2=2.9\,10^{-3}$. 

In summary, we can say that the deviation angle elliptical model can be
applied to NFW mass profile up to $\epsilon\sim0.25$. For this range of values,
$\epsilon$ can be identified with the ellipticity of the potential 
$\epsilon_\varphi$, and the ellipticity of the projected mass density 
$\epsilon_\Sigma$ is about twice larger than $\epsilon$.

\section{Conclusion}

We propose a simple new formalism that introduces the ellipticity into 
the lens potential/deflection-angle of a circular mass model. 
The method can be applied when the lens 
potential or/and the deviation angle takes an analytical form. 
Then for radial mass profiles for which the 2D surface density $\Sigma$ 
also has an analytical expression, this formalism
gives analytical expressions of a pseudo-elliptical mass distribution
for the deviation angle, the projected mass density, the convergence 
and shear.

Whatever the form of the mass distribution, the elliptical parameter 
$\epsilon$ is simply expressed as a function of
the ellipticity of the potential. This is 
particularly helpful in getting some insight on the physical meaning of
this parameter.

We have applied this formalism to the NFW profile and estimated the range
of ellipticity ($\epsilon\lesssim0.25$, or  $\epsilon_\Sigma\lesssim0.4$) for
which this model is a good description of elliptical mass distributions
and thus can be reliably applied to observational data. 
To derive these limits, we introduced a particular fit for elliptical-like 
profiles, that can be useful in similar cases. 

Our proposed  method is particularly useful when it is essential to 
quickly calculate the potential, the deflection angle and magnification of 
many images and/or
many mass clumps. This is particularly important when using {\it inverse}
methods (such as maximum likelihood) to investigate galaxy-galaxy lensing
in the field or in clusters of galaxies, or to compute time delays.

\begin{acknowledgements}
We are grateful to Oliver Czoske, Priya Natarajan, Graham P. Smith and 
Genevi\`eve Soucail 
for useful discussion and a careful reading of this paper. 
We thank the referee Chuck Keeton for interesting remarks, that makes this
formalism even more interesting than we originally thought.
JPK acknowledges CNRS for support. This work
benefits from the LENSNET European Gravitational Lensing Network No.
ER-BFM-RX-CT97-0172.
\end{acknowledgements}

\bibliographystyle{aa}
\bibliography{golse}

\end{document}